\documentclass[12pt]{article}
\usepackage{texdraw}
\usepackage{color}
\usepackage{colordvi}

\begin{document}

\begin{center}

{\Large \bf Gauge Theories for Target Spaces with Degenerate Metrics\footnote{Talk given at 5th International Conference Bolyai-Gauss-Lobachevsky: Methods of Non-Euclidean Geometry in Modern Physics, Minsk, Belarus, October 10--13, 2006 }} 
\end{center}

\begin{center}
{\large \bf N.A. Gromov} \\
 Department of Mathematics, \\
Komi Science Center UrD RAS \\
Kommunisticheskaya st., 24, Syktyvkar, 167982, Russia \\
E-mail: gromov@dm.komisc.ru
\end{center}



 
 \begin{center}
{ \bf Abstract}
\end{center}
 Some gauge theories for fiber target spaces with degenerate metrics are regarded.
The gauge theory with Galilei group $G(2)$ is obtained   
as a contraction of $SO(2)$ gauge theory with Higgs mechanism. The analogue of the standard electroweak theory for contracted $SU(2)$ group is considered. It is shown that the gauge field theory with degenerate metrics in target (matter) field space describe the same set of  fields and particle mass as  initial one, if Lagrangians in the base and   
 in the fiber both are taken into account. 
Such  theory based on non-semisimple contracted group provide more simple field interactions as compared with initial one. The conjecture is advanced  that  Higgs boson being  an artefact of the Higgs mechanism 
is unobservable.

\section{Introduction}
  
 Gauge field theory was suggested by Yang and Mills \cite{YM-54} and is regarded now as most powerfull method for unified description of fundamental interactions in particle physics, where the compact semisimple Lie groups seem to play the most fundamental roles. For example, in the standard Weinberg-Salam model \cite{W-67}, \cite{S-68} of electroweak theory, the gauge group is $SU(2)\times U(1). $ 
   
  It was realized  by Nappi and Witten \cite{NW-93} that one can also construct gauge theories for some non-semisimple groups (which admit a nondegenerate invariant bilinear form ) and such theories have much simpler structure than the standard theories with semisimple groups.
 Later the gauge theories, $\sigma$-models and solitonic hierarchies for  different non-semisimple groups was investigated \cite{Ts-95}--\cite{V-06}.
  
 The aim of this  talk is to regard some gauge theories based on non-semisimple Cayley-Klein groups.
  Such Cayley-Klein groups are invariance groups of spaces with degenerate metrics and can be obtained from the classical simple groups by contractions.
 

\section{Gauge theory  for $SO(2)$ group}
 
   Let $ {\bf R_4}$ is Minkowski space-time: 
    $
    x_\mu x_\mu = x_0^2-x_1^2-x_2^2-x_3^2,  \; \mu=0,1,2,3
    $
  and $\Phi_2$ is the target  space, i.e. the space of fundamental representation of $SO(2)$ group, that elements named matter fields depend on $x \in {\bf R_4}.$
  Gauge transformations:
$\phi'(x)=\omega(\alpha(x))\phi(x), \; \omega(\alpha(x)) \in SO(2)$ or
\begin{equation}
\left(
\begin{array}{c}
	\phi'_1(x) \\
	\phi'_2(x)
\end{array} \right)
=\left(
\begin{array}{rl}
\cos \alpha(x) & \sin \alpha(x)\\
-\sin \alpha(x) & \cos \alpha(x)	
\end{array}\right) 
\left(
\begin{array}{c}
	\phi_1(x) \\
	\phi_2(x)
\end{array} \right)
\label{1}
\end{equation}
leave invariant the form  
$\phi^t\phi= \phi_1^2(x)+\phi_2^2(x) $ 
and define Euclid metrics in $\Phi_2.$

   The Lagrangian is written as  \cite{R-99} 
   \begin{equation}
   L=-\frac{1}{4}F_{\mu\nu}F_{\mu\nu} + \frac{1}{2}(D_\mu \phi)^{t}D_\mu \phi+ \frac{\mu^2}{2}\phi^{t}\phi -\frac{\lambda}{4}(\phi^{t}\phi)^2, 
\label{2}
\end{equation}   
   where covariant derivatives are
   \begin{equation}
   D_\mu\phi_1=\partial_\mu\phi_1 + eA_\mu\phi_2, \quad D_\mu\phi_2=\partial_\mu\phi_2 - eA_\mu\phi_1.
   \label{3}
\end{equation}  
  Here $e$ is the coupling constant,  $A_\mu(x) $ is the gauge field and the  
    field tensor is defined in the standard way
  $
   F_{\mu\nu}=\partial_\mu A_\nu - \partial_\nu A_\mu. 
   $ 
   
 Higgs mechanism \cite{H-64} is the method of generation mass for gauge fields.
 A Lagrangian  ground state is such configuration of fields $A_\mu, \phi_1, \phi_2, $ that minimize theirs
 energy.
   There are a set of ground states
      \begin{equation}
      (\phi_1^{\mbox{vac}})^2 + (\phi_2^{\mbox{vac}})^2 = \phi_0^2, \quad A_\mu^{\mbox{vac}}=\partial_\mu\alpha,\quad \phi_0=\frac{\mu}{\sqrt{\lambda}},
      \label{4}
\end{equation}  
 which can be obtained by gauge transformations from
 one of them:
   \begin{equation}
   A_\mu^{\mbox{vac}}=0, \quad \phi^{\mbox{vac}}=\left(
\begin{array}{c}
	\phi_0 \\
	0
\end{array} 
\right), \quad \phi_0=\frac{\mu}{\sqrt{\lambda}},     
   \label{5}
\end{equation}  
  as it is shown on Fig. 1.  
  
\begin{figure}[h]
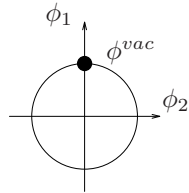

  \centertexdraw{
\drawdim{mm}
\setunitscale 0.5
\bsegment
\move(0 -20)\arrowheadtype t:V \arrowheadsize l:2 w:1 \linewd 0.2 \ravec(0 45)
\move(-20 0)\ravec(40 0)
\textref h:C v:C \htext(24 4) {\footnotesize $\phi_2$}
\textref h:C v:C \htext(-7 27) {\footnotesize $\phi_1$}
\textref h:C v:C \htext(12.0 17) {\footnotesize $\phi^{vac}$}
\move(0 0) \lcir r:14
\move(0 14) \fcir f:0 r:2
\esegment
}
\caption{ Lagrangian  ground states for $SO(2)$ gauge theory.}
\end{figure}
   
    For small (linear) field exitations   with respect of vacuum 
$A_\mu (x),\;\phi_1 (x)=\phi_0 +\chi (x) ,\;\phi_2 (x)  $ 
Lagrangian (\ref{2}) can be written as
 \begin{equation} 
 L=L^{(2)}+L^{(3)}+L^{(4)},
   \label{6}
\end{equation}  
    where quadratic in fields $A_\mu, \chi , \phi_2  $ Lagrangian  
   $$
   L^{(2)}=-{\frac{1}{4}B_{\mu\nu}B_{\mu\nu} + \frac{e^2\phi_0^2}{2}B_\mu B_\mu}
   +{\frac{1}{2}(\partial_\mu\chi)^2 -\mu^2\chi^2},
   $$
   \begin{equation}
   B_\mu = A_\mu - \frac{1}{e\phi_0}\partial_\mu \phi_2, \quad F_{\mu\nu}=B_{\mu\nu}
   \label{7}
\end{equation}  
    describe massive vector field {$B_\mu, \; m_V=e\phi_0=\frac{e\mu}{\sqrt{\lambda}} $} --- 
   gauge field 
      and massive scalar field {$\chi, \; m_\chi=\sqrt{2}\mu $  --- 
    matter field (Higgs boson).
   Field interactions are given  by $L^{(3)}$  and $ L^{(4)}$
   $$
   L^{(3)}=eA_\mu\left(\phi_2\partial_\mu\chi - \chi\partial_\mu\phi_2\right) +\phi_0\chi\left[e^2A_\mu^2 - \lambda\left(\chi^2+\phi_2^2\right)\right],
   $$
   \begin{equation}
   L^{(4)}=\frac{1}{2}\left(\chi^2+\phi_2^2\right)\left[e^2A_\mu^2 - \frac{\lambda}{2}\left(\chi^2+\phi_2^2\right)\right],
   \label{8}
\end{equation}  
which  include terms of third and fourth order in fields.

\section{Gauge theory for Galilei group. }

\subsection{Galilei group  and Galilei geometry }
  Galilei space  $\Phi_2({\iota})$ and Galilei group $G_2=SO(2;{\iota})$ 
   can be obtained from $\Phi_2 $ and $SO(2)$ by substitution:
   $
   \phi_2 \rightarrow {j}\phi_2, \; \alpha \rightarrow {j}\alpha, 
   $ 
   where contraction parameter takes two values ${j}=1,{\iota}, \;\; {\iota^2}=0,\; \iota/\iota=1. $
  Gauge transformations
\begin{equation}
\left(
\begin{array}{c}
	\phi'_1(x) \\
	j\phi'_2(x)
\end{array} \right)
=\left(
\begin{array}{rl}
\cos j\alpha(x) & \sin j\alpha(x)\\
-\sin j\alpha(x) & \cos j\alpha(x)	
\end{array}\right) 
\left(
\begin{array}{c}
	\phi_1(x) \\
	j\phi_2(x)
\end{array} \right)
\label{9}
\end{equation}
leave invariant the form  
$ \phi^t({j})\phi({j})= \phi_1^2+{j^2}\phi_2^2, $ 
which for $ j=1$ define Euclid metrics in $\Phi_2.$
  
For $j={\iota}$  Galilei (degenerate) metrics 
 $
 \phi^t(\iota)\phi(\iota)=\phi_1^2+\iota^2\phi_2^2
 $
in the 2-dim fiber space $\Phi_2(\iota) $
is obtained, where $\{\phi_1\}$ is 1-dim base and $ \{\phi_2\}$ is 1-dim fiber. 
There are {\bf two invariants}: 
$\mbox{inv}_1= \phi_1^2$ under the general transformations
$\phi'({\iota})=\omega({\iota}\alpha)\phi({\iota}), $ where
%
%
%
\begin{equation}   
SO(2;\iota) \ni  \omega({\iota}\alpha)=\left(
\begin{array}{rl}
1 &  {\iota}\alpha \\
- {\iota}\alpha & 1	
\end{array}\right), \; \alpha \in {\bf R}, \quad \omega^t({\iota}\alpha)\omega({\iota}\alpha)=1
\label{10}
\end{equation}
 and $\mbox{inv}_2= \phi_2^2 $ under  transformations in the fiber $(\phi_1=0).$  
   Therefore there are {\bf two metrics}: one in the base and another  in the fiber.

A bundle  of lines through a point on this two  planes has different properties relative to the plane automorphism   \cite{P-65}. 
On  Euclid plane, any two lines of the bundle are transformed to each other by rotation  around the point.
%
  On  Galilei plane, there is one isolated line   that  do not superposed with any other line of the bundle by Galilei boost.
%

If one interpret these planes in some physical context, then on  Euclid plane all lines must have 
the same physical dimension $[\phi_1]=[\phi_2].$
 On Galilei plane, there are infinite many lines with physical dimension identical with  dimension of the base $[\phi_1]$ and one isolated line in the fiber with some different physical dimension 
 $[\phi_2]\neq[\phi_1]\;$ (see \cite{G-06} for details).
 
\subsection{Gauge theory for Galilei group $G_2.$ }

 Gauge theory for  $SO(2;{\iota})=G_2$   can be obtained from thouse for $SO(2)$  by the substitution 
\begin{equation}    
   \phi_1 \rightarrow \phi_1, \quad \phi_2 \rightarrow {j}\phi_2, \quad A_\mu \rightarrow {j}A_\mu, \quad
   F_{\mu\nu} \rightarrow {j}F_{\mu\nu},\quad \mbox{with}\;\; {j=\iota}.
\label{11}
\end{equation}   
  Full Lagrangian is splited  on the Lagrangian in the base
\begin{equation}   
 { L_b=\frac{1}{2}(\partial_\mu\phi_1)^2+\frac{\mu^2}{2}\phi_1^2  -\frac{\lambda}{4}\phi_1^4,}
\label{12}
\end{equation}     
  the  Lagrangian in the fiber $(\approx  {j^2})$
   $$
  {  L_f=-\frac{1}{4}F_{\mu\nu}^2+\frac{1}{2}(\partial_\mu\phi_2)^2+\frac{\mu^2}{2}\phi_2^2+}
   $$
\begin{equation}    
 {  +\frac{1}{2}\phi_1^2 (e^2A_\mu^2 - \lambda\phi_2^2) 
   +eA_\mu\left(\phi_2\partial_\mu\phi_1 - \phi_1\partial_\mu\phi_2\right)},   
\label{13}
\end{equation}       
      and higher order part $(\approx  {j^4}) $ 
\begin{equation}    
  { L_h=\frac{1}{2}\phi_2^2\left( e^2A_\mu^2-\frac{\lambda}{2}\phi_2^2\right)},
\label{14}
\end{equation}    
   which disappear for ${j=\iota}.$
 
 Higgs mechanism  is realized in three steps: 
 
  (i) the Lagrangian  in the base $L_b$ is maximal and the  Lagrangian  in the fiber is equal to zero $L_f=0$ at
\begin{equation}  
  \phi_1=\phi_0, \quad \phi_2=0, \quad A_\mu =\frac{1}{e}\partial_\mu\alpha,\quad F_{\mu\nu}=0,\quad \lambda\phi_0^2=\mu^2,
\label{15}
\end{equation}   
 where point $M(\phi_0,0) \in \Phi_2({\iota})$ is one of the  ground states;
  
(ii) gauge transformations 
\begin{equation}   
  \phi_1'=\phi_1, \quad \phi_2'=\phi_2+\alpha\phi_1, \quad 
  A_\mu' =A_\mu +\frac{1}{e}\partial_\mu\alpha,\quad F_{\mu\nu}'=F_{\mu\nu},
\label{16}
\end{equation}     
  applied to $M$ define the set of ground states $\left\{ \phi_1^2=\phi_0^2,\; \phi_2 \in {\bf R}, \;\; 
 A_\mu' =\frac{1}{e}\partial_\mu\alpha \right\}$ as {sphere}  in   $\Phi_2({\iota})$ (see Fig. 2);
  
\begin{figure}[h]
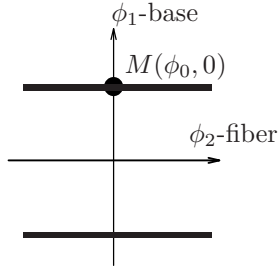

\centertexdraw{
\drawdim{mm}
\setunitscale 0.7
\bsegment
\move(0 -20)\arrowheadtype t:V \arrowheadsize l:2 w:1 \linewd 0.2 \ravec(0 45)
\move(-20 0)\linewd 0.4 \ravec(40 0)
\linewd 0.2
\textref h:C v:C \htext(23 5) {\footnotesize $\phi_2$-fiber}
\textref h:C v:C \htext(8 28) {\footnotesize $\phi_1$-base}
\textref h:C v:C \htext(12 18) {\footnotesize  $M(\phi_0,0)$}
\move(-16 14)\linewd 0.4 \rlvec(32 0)
\move(-16 -14) \rlvec(32 0)
\textref h:C v:C \htext(0 14) { \rule{25mm}{2pt}}
\textref h:C v:C \htext(0 -14) { \rule{25mm}{2pt}}
\move(0 14) \fcir f:0 r:1.8
\esegment
}
\caption{ Ground states for Galilei  gauge theory.}
\end{figure}

  (iii) field excitations arround ground state $M$ are
\begin{equation}    
  \phi_1(x)=\phi_0+\chi(x), \quad \phi_2(x),\quad A_\mu(x).
\label{17}
\end{equation}    
  
%

As a result we obtain 
the  Lagrangian in the base
  $$
  L_b=
{ \frac{1}{2}(\partial_\mu\chi)^2-\mu^2\chi^2 }
  +L_b^{(3)}+ L_b^{(4)},
  $$
\begin{equation}    
 L_b^{(3)}=  -\lambda\phi_0\chi^3, \quad L_b^{(4)}= -\frac{\lambda}{4}\chi^4,
\label{18}
\end{equation}      
  which describe massive scalar  field $\chi, \; m_\chi=\sqrt{2}\mu $ 
  (Higgs boson) and its self interaction $L_b^{(3)},\; L_b^{(4)}$
  and the Lagrangian in the  fiber
   $$
   L_f={-\frac{1}{4}B_{\mu\nu}^2+\frac{e^2\phi_0^2}{2}B_\mu^2} + L_f^{(3)}+L_f^{(4)},
   $$
   $$
  L_f^{(3)}=  eA_\mu\left(\phi_2\partial_\mu\chi - \chi\partial_\mu\phi_2\right) +
  \phi_0\chi \left(e^2A_\mu^2-\lambda\phi_2^2  \right), \quad   
   $$
\begin{equation}    
 L_f^{(4)}= \frac{1}{2}\chi^2 (e^2A_\mu^2 - \frac{\lambda}{2}\phi_2^2),\quad   
   { B_\mu = A_\mu - \frac{1}{e\phi_0}\partial_\mu \phi_2}, 
\label{19}
\end{equation}      
   which describe massive vector  gauge  field $B_\mu, \;   m_V=e\phi_0=\frac{e\mu}{\sqrt{\lambda}} $  
   and field interactions.
  
     So in the  theory with Galilei gauge group $G_2$
    matter field (Higgs boson) ${\chi}$ in the base and gauge field ${B_\mu}$ in the fiber have different physical dimensions. Nevertheless, the  mass dimension   of Higgs boson ${m_\chi=\sqrt{2}\mu} $ and
   vector boson ${ m_V=e\phi_0=\frac{e\mu}{\sqrt{\lambda}} }$ are identical and are the same as for
  $SO(2)$ gauge theory.
  More simple field interactions are provided by  Galilei gauge theory as compared to $SO(2)$ one.


\section{ Electroweak theory and its contraction} 
\subsection{Standard electroweak theory}
  Standard electroweak theory is constructed for gauge group
  $SU(2)\times U(1). $
   Gauge fields $A_\mu^a \; (a=1,2,3)$ correspond to $ SU(2)$ and field $ B_\mu$ correspond to $U(1).$
  There are two coupling constants: $g $ for $SU(2)$ and $g'$ for $U(1).$
The fundamental representation space of $SU(2)$ is interpreted as the target space $\Phi_2({\bf C}). $
   Matter fields $ \phi= \left(
\begin{array}{c}
	\phi_1 \\
	\phi_2
\end{array} \right) \in \Phi_2({\bf C})$
 are now complex ones $\phi_1, \phi_2 \in {\bf C}.$

Standard  bosonic Lagrangian is written in the form \cite{R-99}
\begin{equation}   
  L=-\frac{1}{4}F_{\mu\nu}^aF_{\mu\nu}^a -\frac{1}{4}B_{\mu\nu}B_{\mu\nu}
  + (D_\mu \phi)^{\dagger}D_\mu \phi -\lambda \left(\phi^{\dagger}\phi- \frac{v^2}{2}\right)^2, 
\label{20}
\end{equation}      
where $D_\mu $ is  covariant derivative
   $$
   D_\mu\phi=\partial_\mu\phi -i\frac{g}{2}\tau^a A_\mu^a\phi-i\frac{g'}{2}B_\mu\phi,
   $$
\begin{equation}    
   \tau^1=\left(\begin{array}{cc}
	0 & 1 \\
	1 & 0
\end{array} \right), \quad 
\tau^2=\left(\begin{array}{cc}
	0 & -i \\
	i & 0
\end{array} \right), \quad 
\tau^3=\left(\begin{array}{cc}
	1 & 0 \\
	0 & -1
\end{array} \right). 
\label{21}
\end{equation}    
 Matrices $ T^a=\frac{1}{2}\tau^a$ and $Y=\frac{1}{2} $ are generators of $SU(2)\times U(1).$
One of the Lagrangian  ground states 
\begin{equation}   
  \phi^{vac}=\left(\begin{array}{c}
	\frac{v}{\sqrt{2}}  \\
	0 
\end{array} \right), \quad  A_\mu^a=B_\mu=0
\label{22}
\end{equation}    
is taken as the vacuum. The matrix  
  $
  Q=Y-T^3=\left(\begin{array}{cc}
	0 & 0 \\
	0 & 1
\end{array} \right)
  $
  is generator of $U(1)_{em} $ subgroup, which annihilate ground state $Q\phi^{vac}=0.$
  Linear field exitations with respect of vacuum
\begin{equation}   
  \phi_1(x)=\frac{1}{\sqrt{2}}(v+\chi(x)), \quad \frac{1}{\sqrt{2}}\phi_2(x), \quad A_\mu^a(x), \quad B_\mu(x)
\label{23}
\end{equation}    
are regarded and new fields are introduced 
  $$
  {W_\mu^{\pm}=\frac{1}{\sqrt{2}}\left(A_\mu^1\mp iA_\mu^2  \right)}, 
  $$
\begin{equation}   
 { Z_\mu =\frac{1}{\sqrt{g^2+g'^2}}\left( gA_\mu^3-g'B_\mu \right)},\quad
 { A_\mu =\frac{1}{\sqrt{g^2+g'^2}}\left( gA_\mu^3+g'B_\mu \right)}, 
\label{24}
\end{equation}    
where $W_\mu^{\pm} $ are the complex fields $(W_\mu^{-})^*=W_\mu^{+} $ and $Z_\mu, A_\mu $ are the real fields.

   The second order Lagrangian
  $$
  L^{(2)}=-{\frac{1}{2}W_{\mu\nu}^{+}W_{\mu\nu}^{-}+m_W^2W_\mu^{+}W_\mu^{-} } -{\frac{1}{4}F_{\mu\nu}F_{\mu\nu}}-
  $$
\begin{equation}   
   -{\frac{1}{4}Z_{\mu\nu}Z_{\mu\nu}+\frac{1}{2}m_Z^2Z_\mu Z_\mu} +  
  {\frac{1}{2}\left(\partial_\mu\chi \right)^2 -\frac{1}{2}m_{\chi}^2\chi^2}
\label{25}
\end{equation}      
 describe massive vector fields {$W_\mu^{\pm}, \;  m_W=\frac{1}{2}gv$} ($W$-bosons),
   massless vector field {$A_\mu, \; m_{A}=0$} (photon),
  massive vector field {$Z_\mu, \; m_Z=\frac{v}{2}\sqrt{g^2+g'^2}$} ($Z$-boson)
and  massive scalar field {$ \chi,\; m_{\chi}=\sqrt{2\lambda}v$} (Higgs boson).
$W$- and $Z$-bosons are connected with the weak interaction, photon -- with electromagnetic one. 
  
  $W$- and $Z$-bosons was experimentally observed:
  $
  m_W=80 GeV, \quad m_Z=91 GeV.
  $
  Higgs boson is unobserved up to now.
  $W_\mu^{\pm}, Z_\mu, A_\mu $ are the gauge fields, $ \chi $ is the matter field. 
  Higgs boson is arised in gauge theory with Higgs mechanism for any gauge group.
 Taking into account these arguments we advance the following 
  
  { \bf Conjecture:}
  
  {\it  Higgs boson is an artefact of the Higgs mechanism and therefore is unobservable.}
  
%

\subsection{Contraction of the electroweak theory to $SU(2;\iota)\times U(1)$ group }

 Compact simple group $SU(2)$  is defined as transformation group of $\Phi_2({\bf C})\equiv {\bf C}_2$
 \begin{equation} 
 \omega\phi=\left(
\begin{array}{lr}
\alpha & -\beta^*\\
\beta &	\alpha^*
\end{array}\right)
\left(
\begin{array}{l}
\phi_1 \\
\phi_2
\end{array}\right), \quad \omega \in SU(2), \;\; \phi_1,\phi_2 \in {\bf C},
\label{26}
\end{equation}   
 which leave invariant the form $\phi^{\dagger}\phi=|\phi_1|^2+|\phi_2|^2.$ 
 Here $|\alpha|^2+|\beta|^2=1.$

  Non-compact non-semisimple contracted group $SU(2;{\iota})$  and fiber complex space $\Phi_2({\iota})$ 
   can be obtained from $SU(2)$ and $\Phi_2 $ by the substitution:
 \begin{equation}    
   \phi_2 \rightarrow {j}\phi_2, \quad \beta \rightarrow {j}\beta, \quad {j=\iota},
\label{27}
\end{equation}    
    then $|\alpha|^2+{\iota^2}|\beta|^2=|\alpha|^2=1,$ i.e. 
\begin{equation}    
 SU(2;{\iota})  \ni \omega({\iota})=\left(
\begin{array}{lr}
e^{i\psi}& -{\iota}\beta^*\\
{\iota}\beta &	e^{-i\psi}
\end{array}\right), \; \psi \in [0,2\pi), \; \beta \in {\bf C}. 
\label{28}
\end{equation}     
The fundamental representation of $SU(2;{\iota}) $ group is unitary $ \omega^{\dagger}({\iota})\omega({\iota})=1.$
The space   $\Phi_2({\iota}) $ is  2-dim complex fiber space with degenerate metrics 
\begin{equation}   
 \phi^{\dagger}({\iota})\phi({\iota})={|\phi_1|^2}+{\iota^2|\phi_2|^2},
\label{29}
\end{equation}   
 where 
$\{\phi_1\}$ is the complex base and $ \{\phi_2\}$ is the complex fiber. 
There are {\bf two invariants}: 
$\mbox{inv}_1= |\phi_1|^2$ under the general transformations
\begin{equation}   
 \phi_1'=e^{i\psi}\phi_1,\;\;\phi_2'=e^{-i\psi}\phi_2+\beta\phi_1  
\label{30}
\end{equation}    
 and {$\mbox{inv}_2= |\phi_2|^2 $} under the transformations in the fiber 
 $\phi_1=0, \; \phi_2'=e^{-i\psi}\phi_2.$
   Therefore there are {\bf two metrics}: one in the base and another in the fiber.
 
  Contraction of the standard electroweak theory to $SU(2;{\iota})\times U(1)$ gauge group is obtained by the substitution:
\begin{equation}  
  \phi_1 \rightarrow \phi_1, \;\; \phi_2 \rightarrow {j}\phi_2, \;\; A_\mu^3 \rightarrow A_\mu^3,\;\; 
  A_\mu^1 \rightarrow {j}A_\mu^1,\;\; A_\mu^2 \rightarrow {j}A_\mu^2,\;
\label{31}
\end{equation}     
 or for new fields:
  $$
  \chi \rightarrow \chi, \;\; A_\mu \rightarrow A_\mu \Rightarrow F_{\mu\nu} \rightarrow F_{\mu\nu}, \;\; 
  Z_\mu \rightarrow Z_\mu \Rightarrow Z_{\mu\nu} \rightarrow Z_{\mu\nu},\;\;
  $$
\begin{equation}  
  W_\mu \rightarrow {j}W_\mu \Rightarrow W_{\mu\nu}^{\pm} \rightarrow {j}W_{\mu\nu}^{\pm}.
\label{32}
\end{equation}     
%
  The second order Lagrangian
  $$
  L^{(2)}({j})=
   -{\frac{1}{4}F_{\mu\nu}F_{\mu\nu}}-{\frac{1}{4}Z_{\mu\nu}Z_{\mu\nu}+\frac{1}{2}m_Z^2Z_\mu Z_\mu} +{\frac{1}{2}\left(\partial_\mu\chi \right)^2 -\frac{1}{2}m_{\chi}^2\chi^2}+
  $$
\begin{equation}  
   + {j^2}\left[  -{\frac{1}{2}W_{\mu\nu}^{+}W_{\mu\nu}^{-}+m_W^2W_\mu^{+}W_\mu^{-}}`	 \right]
\label{33}
\end{equation}     
  for $ {j}= {\iota}$ is splited on the Lagrangian in the base
\begin{equation}  
  L^{(2)}_b=
   -{\frac{1}{4}F_{\mu\nu}F_{\mu\nu}} - {\frac{1}{4}Z_{\mu\nu}Z_{\mu\nu}+\frac{1}{2}m_Z^2Z_\mu Z_\mu} +{\frac{1}{2}\left(\partial_\mu\chi \right)^2 -\frac{1}{2}m_{\chi}^2\chi^2},
\label{34}
\end{equation}     
 which describe the gauge fields {$A_\mu,$}  {$Z_\mu, $} as well as the matter field
 {$\chi $} (Higgs boson)
  and the Lagrangian in the fiber $( \approx {\iota^2})$
\begin{equation}  
  L^{(2)}_f=   -{\frac{1}{2}W_{\mu\nu}^{+}W_{\mu\nu}^{-}+m_W^2W_\mu^{+}W_\mu^{-},}
\label{35}
\end{equation}     
  which describe the gauge fields {$ W_\mu^{\pm}. $}
   As regards the field interactions,
   then,  for example, in the interactions of all fields with electromagnetic field 
   $$
   -\frac{1}{4}F^a_{\mu\nu}F^a_{\mu\nu}=-\frac{1}{4}\left(F_{\mu\nu}\sin \theta_w +Z_{\mu\nu}\cos \theta_w  \right)^2 -
   $$
   $$
   -{j^2} \frac{1}{2}\left[| D_\mu W_\nu^{-}+ig\left(Z_\mu W_\nu^{-} - Z_\nu W_\mu^{-}\right)|^2 + \right.
   $$
   $$
  \left. + ig\left( F_{\mu\nu}\sin \theta_w +Z_{\mu\nu}\cos \theta_w \right) 
   \left(W_\mu^{-}W_\nu^{+} -W_\mu^{+}W_\nu^{-}  \right) \right]+ 
   $$
\begin{equation}   
   +{j^4}\frac{1}{4}g^2\left(W_\mu^{-}W_\nu^{+} -W_\mu^{+}W_\nu^{-}  \right)^2,
\label{36}
\end{equation}      
  the terms $( \approx {j^4} )$ disappear under contraction ${j=\iota}, $ 
  whereas those  $( \approx {j^2} )$ are included in $L_f.$
 
 
 Similar to the  Galilei gauge theory, in the  contracted electroweak theory
  the same  fields    with the same particle  mass spectrum as for standard  theory  
 are reproduced but more simple field interactions are provided. 

\section{Conclusion}

 Single contracted Cayley-Klein group is  the motion group  of its   fundamental representation space,
 which has two metrics: one in the base and another in the fiber \cite{G-90}. This means that for the 
 complete description of a physical system in such space
  it is necessary to regard two Lagrangians \cite{G-06}.
    The gauge field theory with degenerate metrics in target (matter) field space describe  the same set of  fields and particle mass as  initial one, if Lagrangians in the  base and  in the  fiber both are taken into account.  
    
     Since it is the structure constants that determine the interactions and since under contractions of Lie group some  structure constants of its algebra turn to zero, the gauge field theory based on non-semisimple  contracted group  provide  more simple field interactions as compared with initial one.  
  
    The experimental  observation of the  gauge field particles predicted by the standard electroweak theory  and  unobservation of the single matter field as well as the presence of Higgs boson in any gauge theory with Higgs mechanism suggests the following 
  
  { \bf Conjecture:}
  
  { \it Higgs boson is an artefact of the Higgs mechanism and therefore is unobservable.}  
 
 This work was partially supported by Russian Foundation for Basic Research under  grant 07-01-00374.

\end{document}